\documentclass[aps,prb,twocolumn,showpacs]{revtex4}
\usepackage{epsfig}
\usepackage{graphicx}
\usepackage{amsfonts}
\usepackage[figuresright]{rotating}
\usepackage{amssymb}
\usepackage{amsmath}
\usepackage{psfrag}
\def\avg#1{\langle#1\rangle}

\def\be{\begin{equation}}
\def\ee{\end{equation}}
\def\bea{\begin{eqnarray}}
\def\eea{\end{eqnarray}}

\def\nn{\nonumber}

\def\tr{\mbox{tr}}

\def\nn{\nonumber}

\begin{document}
\title{Spin-orbit coupled Fermi liquid theory of ultra-cold magnetic dipolar
fermions}
\author{Yi Li}
\affiliation{Department of Physics, University of California, San Diego,
California 92093, USA}
\author{Congjun Wu}
\affiliation{Department of Physics, University of California, San Diego,
California 92093, USA}
\begin{abstract}
We investigate Fermi liquid states of the ultra-cold magnetic dipolar
Fermi gases in the simplest two-component case including both
thermodynamic instabilities and collective excitations.
The magnetic dipolar interaction is invariant under the simultaneous spin-orbit
rotation, but not under either the spin or the orbit one. Therefore,
the corresponding Fermi liquid theory is intrinsically spin-orbit coupled.
This is a fundamental feature of magnetic dipolar Fermi gases
different from electric dipolar ones.
The Landau interaction matrix is calculated and is diagonalized in terms of
the spin-orbit coupled partial-wave channels of the total angular momentum $J$.
The leading thermodynamic instabilities lie in the channels of ferromagnetism
hybridized with the ferronematic order with $J=1^+$ and the spin-current
mode with $J=1^-$, where $+$ and $-$ represent even and odd parities, respectively.
An exotic propagating collective mode is identified as spin-orbit coupled
Fermi surface oscillations in which spin distribution on the Fermi
surface exhibits a topologically nontrivial hedgehog configuration.
\end{abstract}
\maketitle

\section{Introduction}
Recent experimental progress of ultracold electric dipolar heteronuclear
molecules has become a major focus of ultracold atom physics
\cite{ospelkaus2008,ni2008,chotia2011}.
Electric dipole moments are essentially classic polarization vectors
induced by the external electric field.
When they are aligned along the $z$ axis, the electric dipolar interaction
becomes anisotropic exhibiting the $d_{r^2-3z^2}$-type anisotropy.
In Fermi systems, this anisotropy has important effects on many-body physics
including both single-particle and collective properties
\cite{sogo2009,miyakawa2008,fregoso2009,ronen2009,chan2010,
sun2010,yamaguchi2010,Li2010,chang2008,zhang2010,lin2009}.
Fermi surfaces of polarized electric dipolar fermions exhibit quadrupolar
distortion elongated along the $z$ axis \cite{sogo2009,miyakawa2008,zhang2010,
chan2010}.
Various Fermi surface instabilities have been investigated including
the Pomeranchuk type nematic distortions \cite{fregoso2009,chan2010}
and stripelike orderings \cite{sun2010,yamaguchi2010}.
The collective excitations of the zero sound mode exhibit anisotropic
dispersions: The sound velocity is largest if the propagation wavevector
$\vec q$ is along the $z$ axis, and the sound is damped if $\vec q$
lies in the $xy$ plane \cite{chan2010,ronen2009}.
Under the dipolar anisotropy, the phenomenological Landau interaction parameters
become tridiagonal matrices, which are calculated at the Hartree-Fock
level \cite{chan2010,fregoso2009}, and
the anisotropic Fermi liquid theory for such systems has been systematically
studied \cite{chan2010}.

The magnetic dipolar gases are another type of dipolar system.
Compared to the extensive research on electric dipolar Fermi systems,
the study on magnetic dipolar ones is a new direction of research.
On the experimental side, laser cooling and trapping
Fermi atoms with large magnetic dipole moments (e.g., $^{161}$Dy
and $^{163}$Dy with $\mu=10 \mu_B$)\cite{lu2010,youn2010,lu2011}
have been achieved, which provides a new opportunity to study exotic
many-body physics with magnetic dipolar interactions.
There has also been a great amount of progress for realizing Bose-Einstein
condensations of magnetic dipolar atoms \cite{koch2008,lahaye2009,
lahaye2009a,menotti2007,lu2011}.

Although the energy scale of the magnetic dipolar interaction is
much weaker than that of the electric one, it is conceptually more interesting
if magnetic dipoles are not aligned by external fields.
Magnetic dipole moments are proportional to the hyperfine spin
up to a Lande factor, thus, they are quantum-mechanical operators rather than
the nonquantized classic vectors as electric dipole moments are.
Furthermore, there is no need to use external fields to induce magnetic
dipole moments.
In fact, the unpolarized magnetic dipolar systems are isotropic.
The dipolar interaction does not conserve spin nor orbit
angular momentum, but is invariant under simultaneous spin-orbit (SO) rotation.
This is essentially a spin-orbit coupled interaction.
Different from the usual spin-orbit coupling of electrons in solids,
this coupling appears at the interaction level but not
at the kinetic-energy level.

The study of many-body physics of magnetic dipolar Fermi gases is
just at the beginning.
For the Fermi liquid properties,
although magnetic dipolar Fermi gases were studied early in Refs.
[\onlinecite{quintanilla2009}] and [\onlinecite{fregoso2009}], the magnetic dipoles
are frozen, thus, their behavior is not much different from the
electric ones.
It is the spin-orbit coupled nature that distinguishes non-polarized
magnetic dipolar Fermi gases from polarized electric ones.
The study along this line was was pioneered by Fregoso and Fradkin \cite{fregoso2010,fregoso2009a}.
They studied the coupling between ferromagnetic and ferronematic orders,
thus, spin polarization distorts the spherical Fermi surfaces and
leads to a spin-orbit coupling in the single-particle spectrum.

Since Cooper pairing superfluidity is another important aspect
of the many-body phase, we also briefly summarize the current progress in
electric and magnetic dipolar systems.
For the single-component electric dipolar gases, the
simplest possible pairing lies in the
$p$-wave channel because $s$-wave pairing is not allowed by the Pauli
exclusion principle.
The dipolar anisotropy selects the $p_z$-channel pairing
\cite{baranov2002,baranov2004,
baranov2008a,you1999,bruun2008,levinsen2011,potter2010,lutchyn2010}.
Interestingly, for the two-component case, the dipolar interaction
still favors the triplet pairing in the $p_z$ channel even though the
$s$ wave is also allowed.
It provides a robust mechanism for the triplet pairing to the first order in
the interaction strength \cite{samokhin2006,wu2010,shi2009,kain2011}.
The mixing between the singlet and the triplet pairings is with a relative
phase $\pm \frac{\pi}{2}$, which leads to a novel time-reversal
symmetry-breaking pairing state \cite{wu2010}.
The investigation of the unconventional Cooper pairing symmetry in
magnetic dipolar systems was studied by the authors \cite{li2010}.
We have found that it provides a robust mechanism for a novel
$p$-wave ($L=1$) spin triplet ($S=1$) Cooper pairing
to the first order in interaction strength.
It comes directly from the attractive part of the magnetic dipolar interaction.
In comparison, the triplet Cooper pairings in $^3$He and solid-state
systems come from spin fluctuations, which is a second-order effect
in interaction strength \cite{leggett1975,volovik2009}.
Furthermore, that pairing symmetry was not studied in $^3$He systems
before in which orbital and spin angular momenta of the Cooper pair
are entangled into the total angular momentum $J=1$.
In contrast, in the $^3$He-$B$ phase \cite{balian1963}, $L$ and $S$
are combined
as $J=0$, and in the  $^3$He-$A$ phase, $L$ and $S$ are decoupled
and $J$ is not well-defined \cite{brinkman1974,anderson1961}.

Fermi liquid theory is one of the most important paradigms in condensed matter
physics on interacting fermions \cite{negele1988,leggett1975}.
Despite the pioneering papers \cite{quintanilla2009,fregoso2009,
fregoso2010,fregoso2009a},
a systematic study of the Fermi liquid properties of magnetic
dipolar fermions is still lacking in the literature.
In particular,  Landau interaction matrices have not been calculated,
and a systematic analysis of the renormalizations from magnetic dipolar
interactions to thermodynamic quantities has not been performed.
Moreover, collective excitations in magnetic dipolar ultracold fermions have not
been studied before.
All these are essential parts of Fermi liquid theory.
The experimental systems of $^{161}$Dy and $^{163}$Dy are with a very large
hyperfine spin of $F=\frac{21}{2}$, thus the Fermi liquid theory taking
into account of all the complicated spin structure should be very challenging.
We take the first step by considering the simplest case of spin-$\frac{1}{2}$
magnetic dipolar fermions which preserve the essential
features of spin-orbit physics and address the above questions.

In this paper, we systematically investigate the Fermi liquid theory
of the magnetic dipolar systems including both the thermodynamic
properties and the collective excitations, focusing on the spin-orbit
coupled effect.
The Landau interaction functions are calculated and are diagonalized in the
spin-orbit coupled basis.
Renormalizations for thermodynamic quantities and the Pomeranchuk-type
Fermi surface instabilities are studied.
Furthermore, the collective modes are also spin-orbit coupled with a
topologically non-trivial configuration of the spin distribution in
momentum space.
Their dispersion relation and configurations are analyzed.

Upon the completion of this paper, we became aware of the nice work by
Sogo {\it et al.} \cite{sogo2011}.
Reference \onlinecite{sogo2011} constructed the Landau interaction matrix
for dipolar fermions with a general value of spin.
The Pomeranchuk instabilities were analyzed for the special case
of spin $\frac{1}{2}$, and collective excitations were discussed.
Our paper has some overlaps on the above topics with
Ref. [\onlinecite{sogo2011}] but with a significant difference,
including the physical interpretation
of the Pomeranchuk instability in the $J=1^-$ channel and our discovery
of an exotic propagating spin-orbit sound mode.

The remaining part of this paper is organized as follows.
The magnetic dipolar interaction is introduced in Sec. \ref{sect:ham}.
The Landau interaction matrix is constructed at the Hartree-Fock
level and is diagonalized in Sec. \ref{sect:landau}.
In Sec. \ref{sect:thermo}, we present the study of the
Fermi liquid renormalization to thermodynamic properties from
the magnetic dipolar interaction.
The leading Pomeranchuk instabilities are analyzed.
In Sec. \ref{sect:coll}, the spin-orbit coupled Boltzmann
equation is constructed.
We further perform the calculation of propagating spin-orbit
coupled collective modes.
We summarize the paper in Sec. \ref{sect:conc}.

\section{Magnetic Dipolar Hamiltonian}
\label{sect:ham}
We introduce the magnetic dipolar interaction and the subtlety of
its Fourier transform in this section.

The magnetic dipolar interaction between two spin-$\frac{1}{2}$ particles
located at $\vec r_{1,2}$ reads
\bea
V_{\alpha \beta; \beta^{\prime}\alpha^{\prime} }(\vec r) =\frac{\mu ^{2}}{r^{3}}
\left[ \vec S_{\alpha \alpha^{\prime}}\cdot \vec S_{\beta \beta^{\prime}}
-3( \vec S_{\alpha \alpha^{\prime}}\cdot \hat{r})
( \vec S_{\beta \beta^{\prime}}\cdot \hat{r}) \right],
\label{eq:mgdpr_r}
\eea
where $\vec S= \frac{1}{2}\vec \sigma$; $\alpha,\alpha^{\prime},
\beta, \beta^{\prime}$ take values of $\uparrow$ and $\downarrow$;
$\vec r=\vec r_1 - \vec r_2$ and $\hat r=\vec r /r$ is
the unit vector along $\vec r$.

The Fourier transform of Eq. (\ref{eq:mgdpr_r}) is
\bea
V_{\alpha \beta; \beta^{\prime}\alpha^{\prime} }(\vec q) =\frac{4 \pi \mu ^{2}}{3}
\left[ 3( \vec {S}_{\alpha \alpha^{\prime}}\cdot \hat{q})
( \vec {S}_{\beta \beta^{\prime}}\cdot \hat{q})
-\vec{S}_{\alpha \alpha^{\prime}}\cdot \vec{S}_{\beta \beta^{\prime}}
\right],
\label{eq:fourier_polar}
\eea
which depends on the direction along the momentum transfer
but not its magnitude.
It is singular as $\vec q \rightarrow 0$.
More rigorously, $V_{\alpha\beta,\beta^\prime\alpha^\prime}(\vec q)$ should be
further multiplied by a numeric factor \cite{chan2010} as
\bea
g(q)=3 \Big(\frac{j_1(q \epsilon)}{q\epsilon}-
\frac{j_1(q L)}{qL} \Big),
\label{eq:form_factor}
\eea
where $\epsilon$ is a short range scale cut off, and $L$ is the long
distance cut off at the scale of  sample size.
The spherical Bessel function $j_1(x)$ shows the asymptotic behavior $j_1(x)\rightarrow \frac{x}{3}$ at $x\rightarrow 0$,
and $j_1(x)\rightarrow \frac{1}{x} \sin (x-\frac{\pi}{2})$
as $x\rightarrow \infty$.
In the long wavelength limit satisfying $q\epsilon\rightarrow 0$ and
$q L\rightarrow \infty$, $g(q)\rightarrow 1$ and we recover
Eq. (\ref{eq:fourier_polar}).
If $\vec q$ is exactly zero,
$V_{\alpha\beta;\beta^\prime \alpha^\prime}=0$, because
the dipolar interaction is neither purely repulsive nor attractive,
and its spatial average is zero.

The second quantization form for the magnetic dipolar interaction is
expressed as
\bea
H_{int}&=&\frac{1}{2V}\sum_{\vec k,\vec k^\prime,\vec q}
\psi^\dagger_\alpha(\vec k+\vec q) \psi^\dagger_\beta(\vec k^\prime)
V_{\alpha\beta;\beta^\prime\alpha^\prime}(\vec q)\nn \\
&\times& \psi_{\beta^\prime}(\vec k^\prime+\vec q) \psi_{\alpha^\prime}(\vec k),
\eea
where $V$ is the volume of the system.
The density of states of two-component Fermi gases at the Fermi energy
is $N_0=\frac{mk_f}{\pi^2\hbar^2}$, and we define a dimensionless
parameter $\lambda= N_0 \mu^2$.
$\lambda$ describes the interaction strength, which equals the
ratio between the average interaction energy and the Fermi energy
up to a factor on the order of $1$.


\section{Spin-orbit coupled Landau interaction }
\label{sect:landau}
In this section, we present the Landau interaction functions of
the magnetic dipolar Fermi liquid, and perform the spin-orbit
coupled partial wave decomposition.

\subsection{The Landau interaction function}
Interaction effects in the Fermi liquid theory are captured by the Landau
interaction function.
It describes the particle-hole channel forward-scattering amplitudes
among quasiparticles on the Fermi surface.
At the Hartree-Fock level, the Landau function is expressed as
\bea
f_{\alpha\alpha^\prime,\beta\beta^\prime}(\hat k, \hat k^\prime)
=f^{H}_{\alpha\alpha^\prime,\beta\beta^\prime}(\hat q)
+f^{F}_{\alpha\alpha^\prime,\beta\beta^\prime}(\hat k,\hat k^\prime),
\eea
where $\vec k$ and $\vec k^\prime$ are at the Fermi surface
with the magnitude of $k_f$
and $\vec q$ is the small momentum transfer in the forward scattering
process in the particle-hole channel.
$f^{H}_{\alpha\alpha^\prime,\beta\beta^\prime}(\vec q)=
V_{\alpha\beta,\beta^\prime\alpha^\prime}(\hat q)$ is the direct
Hartree interaction, and
$f^{F}_{\alpha\alpha^\prime,\beta\beta^\prime}(\vec k; \vec k^\prime)=
- V_{\alpha\beta,\alpha^\prime\beta^\prime} (\vec k-\vec k^\prime)$
is the exchange Fock interaction.
As $\vec q \rightarrow 0$, $f^{H}$ is singular, thus we need to keep
its dependence on the direction of $\hat q$.
More explicitly,
\bea
f^{H}_{\alpha\alpha^\prime,\beta\beta^\prime}(\hat q)&=&
\frac{\pi \mu ^{2}}{3} M_{\alpha\alpha^\prime,\beta\beta^\prime} (\hat q),
\label{eq:f_H}
\\
f^{F}_{\alpha\alpha^\prime,\beta\beta^\prime}(\hat k;\hat k^\prime)&=&
-\frac{\pi \mu ^{2}}{3} M_{\alpha\alpha^\prime,\beta\beta^\prime} (\hat m),
\label{eq:f_F}
\eea
where the tensor is defined as
$M_{\alpha\alpha^\prime,\beta\beta^\prime} (\hat q)
=3(\vec \sigma_{\alpha\alpha^\prime} \cdot \hat q)(\vec
\sigma_{\beta\beta^\prime} \cdot \hat q)
-\vec \sigma_{\alpha\alpha^\prime} \cdot \vec \sigma_{\beta\beta^\prime}$
and $\hat m$ is the unit vector along the direction of the
momentum transfer $\hat m= \frac{\vec k -\vec k^\prime}{
|\vec k -\vec k^\prime|}$.
We have used the following identity:
\bea
&&3(\vec \sigma_{\alpha\beta^\prime} \cdot \hat m)
(\vec \sigma_{\beta\alpha^\prime} \cdot \hat m)
-\vec \sigma_{\alpha\beta^\prime}\cdot \vec \sigma_{\beta\alpha^\prime}\nn \\
&=&3(\vec \sigma_{\alpha\alpha^\prime} \cdot \hat m)
(\vec \sigma_{\beta\beta^\prime} \cdot \hat m)
-\vec \sigma_{\alpha\alpha^\prime} \cdot \vec \sigma_{\beta\beta^\prime}
\eea
to obtain Eq. (\ref{eq:f_F}).

\subsection{The spin-orbit coupled basis}
\label{sect:basis}

Due to the spin-orbit nature of the magnetic dipolar interaction,
we introduce the spin-orbit coupled partial-wave basis for the
quasiparticle distribution over the Fermi surface following
the steps below.

The $\delta n_{\alpha\alpha^\prime}(\vec k)$ is defined as
\bea
\delta n_{\alpha\alpha^\prime}(\vec k)= n_{\alpha\alpha^\prime}(\vec k)
-\delta_{\alpha\alpha^\prime}n_0(\vec k),
\eea
where $n_{\alpha\alpha^\prime}(\vec k)=\avg{\psi^\dagger_{\alpha} (\vec k)
\psi_{\alpha^\prime}(\vec k)}$ is the Hermitian single-particle density
matrix with momentum $\vec k$ and satisfies $n_{\alpha\alpha^\prime}
=n^*_{\alpha^\prime \alpha}$ and $n_0(\vec k)$ is the zero-temperature
equilibrium Fermi distribution function $n_0(\vec k)=1-\theta(k-k_f)$.
$\delta n_{\alpha\alpha^\prime}(\vec k)$ is expanded in terms of the
particle-hole angular momentum basis as
\bea
\delta n_{\alpha\alpha^\prime}(\vec k)&=&\sum_{Ss_z} \delta n_{Ss_z}(\vec k)
\chi_{Ss_z,\alpha\alpha^\prime} \nn \\
&=& \sum_{Ss_z} \delta n_{Ss_z}^*(\vec k) \chi^\dagger_{Ss_z,\alpha\alpha^\prime},
\label{eq:p_h_basis}
\eea
where $\chi_{Ss_z,\alpha\alpha^\prime}$ are the bases for the
particle-hole singlet (density) channel with $S=0$
and triplet (spin) channel with $S=1$, respectively.
They are defined as
\bea
\chi_{00,\alpha\alpha^\prime}&=&\delta_{\alpha\alpha^\prime}, \nn \\
\chi_{10,\alpha\alpha^\prime}&=& \sigma_{z,\alpha\alpha^\prime}, \ \ \,
\chi_{1\pm 1,\alpha\alpha^\prime}=\frac{\mp1}{\sqrt 2}
(\sigma_{x,\alpha\alpha^\prime} \pm i\sigma_{y,\alpha\alpha^\prime}), \nn\\
\eea
which satisfy the orthonormal condition
$\tr(\chi^\dagger_{Ss_z}\chi_{S^\prime s^\prime_z})=2\delta_{SS^\prime}\delta_{s_zs_z^\prime}$.

Since quasiparticles are only well defined around the Fermi surface,
we integrate out the radial direction and arrive at the angular
distribution,
\bea
\delta n_{\alpha\alpha^\prime}(\hat k) =\int \frac{k^2 dk }{(2\pi)^3}
\delta n_{\alpha\alpha^\prime}(\vec k).
\label{eq:ang_distr}
\eea
Please note that angular integration is not performed in Eq. (\ref{eq:ang_distr}).
We expand $\delta n_{\alpha\alpha^\prime}(\hat k)$ in the spin-orbit
decoupled bases as
\bea
\delta n_{\alpha\alpha^\prime}(\hat k)
&=&\sum_{LmSs_z} \delta n_{LmSs_z} Y_{Lm}(\hat k)
\chi_{Ss_z,\alpha\alpha^\prime}, \nn \\
&=& \sum_{LmSs_z} \delta n^*_{LmSs_z} Y^*_{Lm}(\hat k)
\chi^\dagger_{Ss_z,\alpha\alpha^\prime},
\eea
where $Y_{Lm}(\hat k)$ is the spherical harmonics satisfying the normalization
condition $\int d \hat k Y_{Lm}^*(\hat k) Y_{Lm}(\hat k)=1$.

We can also define the spin-orbit coupled basis as
\bea
{\cal Y}_{JJz;LS}(\hat k, \alpha\alpha^\prime)
&=&\sum_{ms_z}\avg{LmSs_z|JJ_z} Y_{Lm}(\hat k)
\chi_{Ss_z,\alpha\alpha^\prime}, \nn \\
{\cal Y}^\dagger_{JJz;LS}(\hat k, \alpha\alpha^\prime)
&=&\sum_{ms_z}\avg{LmSs_z|JJ_z} Y^*_{Lm}(\hat k)
\chi^\dagger_{Ss_z,\alpha\alpha^\prime}, \nn \\
\eea
where $\avg{LmSs_z|JJ_z}$ is the Clebsch-Gordon coefficient
and ${\cal Y}_{JJz;LS}$ satisfies the orthonormal condition of
\bea
&&\int d\hat k ~\tr [
{\cal Y}^\dagger_{JJ_z;LS} (\hat k)
{\cal Y}_{J^\prime J_z^\prime;L^\prime S^\prime } (\hat k)]
=2\delta_{JJ^\prime}\delta_{J_zJ_z^\prime} \delta_{LL^\prime} \delta_{SS^\prime}.
\nn \\
\eea
Using the spin-orbit coupled basis, $\delta n_{\alpha\alpha^\prime}(\hat k)$
is expanded as
\bea
\delta n_{\alpha\alpha^\prime}(\hat k)
&=&\sum_{JJ_z;LS} \delta n_{JJ_z;LS} ~
{\cal Y}_{JJ_z;LS}(\hat k, \alpha\alpha^\prime)\nn \\
&=&\sum_{JJ_z;LS} \delta n^*_{JJ_z;LS} {\cal Y}^{\dagger}_{JJ_z;LS}(\hat k,
\alpha\alpha^\prime),
\eea
where
$\delta n_{JJz;LS}=\sum_{ms_z} \avg{LmSs_z|JJ_z}
\delta n_{LmSs_z}$.

\subsection{Partial-wave decomposition of the Landau function}

We are ready to perform the partial-wave decomposition for Landau interaction
functions.
The tensor structures in Eqs. (\ref{eq:f_H}) and (\ref{eq:f_F})
only depend on $\vec \sigma_{\alpha\alpha^\prime}$ and
$\vec \sigma_{\beta\beta^\prime}$, thus
the magnetic dipolar interaction only contributes to the
spin-channel Landau parameters, i.e., $S=1$.
In the spin-orbit decoupled basis, the Landau functions
of the Hartree and Fock channels are expanded, respectively, as
\bea
\frac{N_0}{4\pi} f^{H,F}_{\alpha\alpha^\prime;\beta\beta^\prime}(\hat k, \hat k^\prime)
&=&\sum_{Lms_z;L^\prime m^\prime  s_z^\prime}
Y_{Lm}(\hat k) \chi_{1s_z}(\alpha\alpha^\prime)\nn \\
&\times& T^{H,F}_{Lm1s_z;L^\prime m^\prime 1 s_z^\prime}
Y^*_{L^\prime m^\prime}(\hat k^\prime ) \chi^\dagger_{1s_z^\prime}(\beta\beta^\prime).
\nn \\
\label{eq:so_decouple_basis}
\eea
For later convenience, we have multiplied the density of states $N_0$ and the
factor of $1/4\pi$ such that $T^{H,F}$  are dimensionless matrices.
Without loss of generality, in the Hartree channel,
we choose $\hat q=\hat z$.

The matrix elements in Eq. (\ref{eq:so_decouple_basis}) are presented below.
In the Hartree channel,
\begin{widetext}
\bea
T^{H}_{Lm1s_z;L^\prime m^\prime 1 s^\prime_z}&=&
\frac{\pi \lambda}{3} (2 \delta_{s_z,0}-\delta_{s_z, \pm 1})
 \delta_{L,0}\delta_{L^\prime, 0}
\delta_{m,0} \delta_{m^\prime,0} \delta_{s_zs_z^\prime};
\eea
and in the Fock channel,
\bea
T^{F}_{Lm1s_z;L^\prime m^\prime 1s^\prime_z}&=&
-\frac{\pi \lambda}{2}
\Big(\frac{\delta_{LL^\prime}}{L(L+1)}
-\frac{\delta_{L+2,L^\prime}}{3(L+1)(L+2)}
-\frac{\delta_{L-2,L^\prime}}{3(L-1)L}\Big) \nn \\
&\times& \int d \Omega_r [ \delta_{s_zs_z^\prime}
-4\pi Y_{1s_z}(\Omega_r) Y^*_{1s_z^\prime}(\Omega_r)]
Y_{Lm}(\Omega_r) Y_{L^\prime m^\prime}^* (\Omega_r).
\eea
The magnetic dipolar interaction is isotropic, thus the spin-orbit
coupled basis are the most convenient.
In these basis, the Landau matrix is diagonal with
respect to the total angular momentum $J$ and its $z$-component
$J_z$ as
\bea
\frac{N_0}{4\pi } f_{\alpha\alpha^\prime;\beta\beta^\prime}(\hat k, \hat k^\prime)=
\sum_{JJ_z L L^\prime}
{\cal Y}_{JJz;L1}(\hat k,\alpha\alpha^\prime)
 F_{JJ_z L1;JJ_z L^\prime 1} ~
{\cal Y}^\dagger_{JJz;L^\prime 1}(\hat k,\beta\beta^\prime).
\eea
The matrix kernel $F_{JJ_zL1;JJ_zL^\prime1}$ reads as
\bea
F_{JJ_zL1;JJ_zL^\prime 1}&=&\frac{\pi\lambda}{3}\delta_{J,1}\delta_{L,0}
\delta_{L^\prime,0}
(2 \delta_{J_z,0}-\delta_{J_z, \pm1}) +
\sum_{ms_z; m^\prime s_z^\prime} %
\avg{Lm1s_z|JJ_z}
\avg{L^\prime m^\prime 1s_z^\prime|JJ_z}
T^{F}_{Lm1s_z;L^\prime m^\prime 1s^\prime_z}.
\label{eq:LL_para}
\eea
\end{widetext}
We found that up to a positive numeric factor, the second term in
Eq. (\ref{eq:LL_para}) is the same as the partial-wave matrices
in the particle-particle pairing channel, which was derived for the analysis
of the Cooper pairing instability in magnetic dipolar systems
\cite{li2010}.

However, the above matrix kernel $F_{JJ_zL1;JJ_zL^\prime1}$ is not diagonal
for channels with the same values of $J J_z$ but different
orbital angular momentum indices $L$ and $L^\prime$.
Moreover, the conservation of parity requires that even and odd
values of $L$ do not mix.
Consequently, $F_{JJ_zL1;JJ_zL^\prime1}$ is either diagonalized
or reduced into a small size of just $2\times 2$.
For later convenience of studying collective modes and thermodynamic
instabilities, we present below the prominent Landau parameters in some
low partial-wave channels.
Below, we use $(J^{\pm} J_zLS)$ to represent these channels in which
$\pm$ represents even and odd parities, respectively.

The parity odd channel of $J=0^-$ only has one possibility of $(0^-011)$
in which
\bea
F_{0^-011;0^-011}=\frac{\pi}{2} \lambda.
\eea
There is another even parity density channel with $J=0^+$, i.e.,
$(0^+000)$, which receives contribution from short range $s$-wave
interaction but no contribution from the magnetic dipolar interaction
at the Hartree-Fock level.
The parity odd channel of $J=1^-$ only comes from $(1^-J_z 11)$ in which
\bea
F_{1^-J_z11;1^-J_z11}= - \frac{\pi}{4} \lambda.
\label{eq:landau_0}
\eea
Another channel of $J=1^-$, i.e., $(1^-J_z10)$, channel from the $p$-wave
channel density interactions, which again receives no contribution from
magnetic dipolar interaction at the Hartree-Fock level.
These two $J=1^-$ modes are spin- and charge-current modes, respectively,
and thus, do not mix due to their opposite symmetry properties under time-reversal
transformation.

We next consider the even parity channels.
The $J=1^+$ channels include two possibilities of $(JJ_zLS)=(1^+J_z01),
(1^+J_z21)$.
The former is the ferromagnetism channel, and the latter is denoted as
the ferronematic channel in Refs. [\onlinecite{fregoso2009}] and [\onlinecite{fregoso2010}].
Due to the spin-orbit nature of the magnetic dipolar interaction,
these two channels are no longer independent but are coupled to each other.
Because the Hartree term breaks the rotational symmetry,
the hybridization matrices for $J_z=0,\pm 1$ are different.
For the case of $J_z=0$, it is
\bea
F_{1^+0}&=&
\left(
\begin{array}{cc}
F_{1001;1001}& F_{1001;1021}\\
F_{1021;1001} & F_{1021;1021}
\end{array}
\right)=
\frac{\pi \lambda}{12}
\left(
\begin{array}{cc}
8 & \sqrt{2} \\
 \sqrt{2} & 1
\end{array}
\right) ,
\label{eq:landau_1} \nn \\
\eea
whose two eigenvalues and their associated eigenvectors are
\bea
w^{1^+0}_1 &=& 0.69 \pi\lambda, \ \ \, \psi^{1^+0}_1=(0.98,0.19)^T, \nn \\
w^{1^+0}_2 &=& 0.06 \pi \lambda, \ \ \, \psi^{1^+0}_2=(-0.19,0.98)^T.
\label{eq:eigen_1}
\eea
The hybridization is small.
For the case of $J_z=\pm 1$, the Landau matrices are the same as
\bea
F_{1^+1}&=&
\left(
\begin{array}{cc}
F_{1101;1101}& F_{1101;1121}\\
F_{1121;1101}& F_{1121;1121}
\end{array}
\right)
=
\frac{\pi \lambda}{12}
\left(
\begin{array}{cc}
-4 & \sqrt{2} \\
 \sqrt{2} & 1
\end{array}
\right) . \nn \\
\label{eq:landau_2}
\eea
Again the hybridization is small as shown in the
eigenvalues and their associated eigenvectors
\bea
w^{1^+1}_1 &=& -0.37 \pi\lambda, \ \ \, \psi^{1^+ 1}_1=(0.97,-0.25)^T, \nn \\
w^{1^+1}_2 &=&  0.12 \pi \lambda, \ \ \, \psi^{1^+ 1}_2=(0.25,0.97)^T.
\label{eq:eigen_2}
\eea
Landau parameters, or, matrices, in other
high partial-wave channels are neglected, because their magnitudes
are significantly smaller than those above.

We need to be cautious on using Eqs. (\ref{eq:landau_1}) and
(\ref{eq:landau_2}) in which the Hartree contribution of Eq. \ref{eq:f_H}
is taken.
However, Eq. (\ref{eq:f_H}) is valid in the limit $q\ll k_f$
but should be much larger than the inverse of sample size $1/L$.
It is valid to use Eqs. (\ref{eq:landau_1}) and (\ref{eq:landau_2}) when
studying the collective spin excitations in Sec. \ref{sect:coll} below.
However, when studying thermodynamic properties, say, magnetic
susceptibility,  under the external magnetic-field uniform at the
scale of $L$, the induced magnetization is also uniform.
In this case, the Hartree contribution is suppressed to zero, thus
the Landau matrices in the $J=1^+$ channel are the same for
all the values of $J_z$ as
\bea
F_{1^+,thm}(\lambda)&=&
\left(
\begin{array}{cc}
F_{1J_z01;1J_z01}& F_{1J_z01;1J_z21}\\
F_{1J_z21;1J_z01} & F_{1J_z21;1J_z21}
\end{array}
\right)_{thm} \nn \\
&=&
\frac{\pi \lambda}{12}
\left(
\begin{array}{cc}
0 & \sqrt{2} \\
 \sqrt{2} & 1
\end{array}
\right) . \ \ \,
\label{eq:landau_3}
\eea
In this case, the hybridization between these two channels
is quite significant.
The two eigenvalues and their associated eigenvectors are
\bea
w^{1^+}_1 &=& -\frac{\pi}{12}  \lambda, \ \ \,
\psi^{1^+}_1= (\sqrt{\frac{2}{3}}, -\sqrt{\frac{1}{3}})^T,
\nn \\
w^{1^+}_2 &=& \frac{\pi }{6} \lambda, \ \ \, \psi^{1^+}_2=
(\sqrt{\frac{1}{3}}, \sqrt{\frac{2}{3}})^T.
\label{eq:eigen_3}
\eea


\section{Thermodynamic quantities}
\label{sect:thermo}

In this section, we study the renormalizations for thermodynamic
properties by the magnetic dipolar interaction and
investigate the Pomeranchuk-type Fermi surface instabilities.

\subsection{Thermodynamics susceptibilities}
The change in the ground-state energy with respect to the variation in
the Fermi distribution density matrix include the kinetic and
interaction parts as
\bea
\frac{\delta E }{V}=  \frac{\delta E_{kin}}{V} +\frac{\delta E_{int}}{V}.
\eea
The kinetic-energy variation is expressed in terms of the angular
distribution of $\delta n_{\alpha\alpha^\prime}(\hat k)$ as
\bea
\frac{\delta E_{kin} }{V}
&=& \frac{4\pi}{N_0} \sum_{\alpha\alpha^\prime}
\int d \hat k \delta n_{\alpha\alpha^\prime}
(\hat k) \delta n_{\alpha^\prime\alpha} (\hat k)\nn \\
&=& \frac{8\pi}{N_0} \sum_{LmSS_z} \delta n^*_{LmSs_z}  \delta n_{LmSs_z},
\eea
where the units of $\delta n_{Ss_z}(\hat k)$  and $\delta n_{LmSs_z}$
are the same as the inverse of the volume.
The variation in the interaction energy is
\bea
\frac{\delta E_{int}}{V}&= &\frac{1}{2}
\sum_{\alpha \alpha^\prime \beta \beta^\prime}
\iint d \hat k d\hat k^\prime
f_{\alpha\alpha^\prime,\beta\beta^\prime}(\hat k, \hat k^\prime)
\delta n_{\alpha^\prime\alpha}(\hat k)
\delta n_{\beta^\prime\beta}(\hat k^\prime) \nn \\
&=& 2 \sum_{Lms_zL^\prime m^\prime s_z^\prime;S}
\delta n^*_{Lm S s_z } f_{LmSs_z,L^\prime m^\prime S s_z^\prime}
\delta n^*_{L^\prime m^\prime S s_z^\prime }. \nn \\
\eea
Adding them together and changing to the spin-orbit coupled basis, we arrive at
\bea
\frac{\delta E}{V}=\frac{8\pi}{N_0} \sum_{JJ_z;LL^\prime;S}
\delta n^*_{JJ_z;LS} M_{JJ_zLS; JJ_z L^\prime S} \delta n_{JJ_z;L^\prime S}, \ \ \,
\eea
where the matrix elements are
\bea
M_{JJ_zLS; JJ_z L^\prime S} =\delta_{LL^\prime} + F_{JJ_zLS;JJ_z L^\prime S}.
\label{eq:matrix}
\eea

In the presence of the external field $h_{JJ_zLS}$, the ground state energy
becomes
\bea
\frac{\delta E}{V}
&=&16\pi\Big\{ \frac{1}{2\chi_0} \sum_{JJ_z L L^\prime S}
\delta n^*_{JJ_z;LS} M_{JJ_zLS; JJ_z L^\prime S} \delta n_{JJ_z;L^\prime S} \nn \\
&-& \sum_{JJ_z L S} h_{JJ_zLS} \delta n_{JJ_z;L S}
\Big\},
\eea
where $\chi_0=N_0$ is the Fermi liquid density of states.
At the Hartree-Fock level, $N_0$ receives no renormalization from the
magnetic dipolar interaction.
The expectation value of $\delta n_{JJ_zLS}$ is calculated as
\bea
\delta n_{JJ_zLS}=\chi_0 \sum_{L^\prime}(M)^{-1}_{JJ_zLS;JJ_z L^\prime S}
h_{JJ_zL^\prime S}.
\eea

For the $J=1^+$ channel, $M^{-1}\approx I-F_{1^+,thm}(\lambda)$
up to first order of $\lambda$ in the case of $\lambda\ll 1$.
As a result, the external magnetic field $\vec h$ along the $z$ axis
not only induces the $z$-component spin polarization, but also induces a
spin-nematic order in the channel of $(J^+J_zLS)=(1^+021)$,
which is an effective spin-orbit coupling term as
\bea
\delta H&=& \frac{\sqrt 2 }{12} \pi \lambda h  \sum_{k}
\psi^\dagger_{\alpha}(\vec k) \Big\{ 
\big[(k^2-3k_z^2) \sigma_z  \nn \\
&-& 3 k_z (k_x \sigma_x + k_y \sigma_y) \big] \Big\}
\psi_\beta (\vec k).
\label{eq:so_1}
\eea
Apparently, this term breaks time-reversal symmetry, and thus cannot
be induced by the relativistic spin-orbit coupling in solid states.
This magnetic field induced spin-orbit coupling in magnetic
dipolar systems was studied by Fregoso {\it et al.}
\cite{fregoso2009,fregoso2010}

\subsection{Pomeranchuk instabilities}
Even in the absence of external fields, Fermi surfaces can be distorted
spontaneously known as Pomeranchuk instabilities \cite{pomeranchuk1959}.
Intuitively, we can imagine the Fermi surface as the elastic membrane
in momentum space.
The instabilities occur if the surface tension in any of its
partial-wave channels becomes negative.
In the magnetic dipolar Fermi liquid, the thermodynamic stability condition
is equivalent to the fact that all the eigenvalues of the matrix $M_{JJ_zLS;JJ_zL^\prime S}$
are positive.

We next check the negative eigenvalues of the Landau matrix in each
partial-wave channel.
Due to the absence of external fields, the Pomeranchuk instabilities
are allowed to occur as a density wave state with a long wave length
$q\rightarrow 0$.
For the case of $J=1^+$, it is clear that in the channel of $J_z=\pm 1$,
the eigenvalue $w_1^{1^+1}$ in Eq. (\ref {eq:eigen_2}) is negative and the
largest among all the channels.
Thus the leading channel instability is in the $(JJ_z)=(1^+\pm 1)$ channel,
which occurs at $w_1^{1^+1}<-1$, or, equivalently, $\lambda>\lambda_{1^+1}^c=0.86$.
The corresponding eigenvector shows that it is mostly a ferromagnetism
order parameter with small hybridization with the ferronematic channel.
A repulsive short-range $s$ wave scattering, which we neglected above will
enhance ferromagnetism and, thus, will drive $\lambda_{1^+1}^c$ to a smaller
value.
The wavevector $\vec q$ of the spin polarization should be on the order
of $1/L$ to minimize the energy cost of twisting spin,
thus, essentially exhibiting a domain structure.
The spatial configuration of the spin distribution should be complicated by
actual boundary conditions.
In particular, the three-vector nature of spins implies the rich configurations of
spin textures.
An interesting result is that the external magnetic field actually
weakens the ferromagnetism instability.
If the spin polarization is aligned by the external field, the Landau
interaction matrix changes to Eq. (\ref{eq:landau_3}).
The magnitude of the negative eigenvalue is significantly smaller than
that of Eq. (\ref{eq:landau_2}).
As a result, an infinitesimal external field cannot align the spin
polarization to be uniform but a finite amplitude is needed.

For simplicity, we only consider ferromagnetism with a single plane
wave vector $\vec q$ along the $z$ axis, then the spin polarization
spirals in the $xy$-plane.
Since $ q\sim 1/L$, we
can still treat a uniform spin polarization over a distance large
comparable to the microscopic length scale.
Without loss of generality, we set the spin polarization along the $x$ axis.
As shown in Ref. \onlinecite{fregoso2010}, ferromagnetism induces
ferronematic ordering.
The induced ferronematic ordering is also along
the $x$ axis, whose spin-orbit coupling can be obtained based on
Eq. (\ref{eq:so_1}) by a permutation among components of $\vec k$ as
$H_{so}^\prime (\vec k) \propto (k^2-3k_x^2) \sigma_x -3 k_x
( k_y \sigma_y +k_z \sigma_z) $.
According to Eq. (\ref{eq:eigen_2}), ferromagnetism and ferronematic
orders are not strongly hybridized, the energy scale of the
ferronematic SO coupling is about $1$ order smaller than that of
ferromagnetism.
An interesting point of this ferromagnetism is that it distorts
the spherical shape of the Fermi surface as pointed by Fregoso and Fradkin \cite{fregoso2010}.
This anisotropy will also affect the propagation of Goldstone modes.
Furthermore, spin waves couple to the oscillation of the shape
of Fermi surfaces bringing Landau damping to spin waves.
This may result in non-Fermi liquid behavior for fermion excitations,
and will be studied in a later paper.
This effect in the nematic symmetry-breaking Fermi liquid state
has been extensively studied before in the literature
\cite{oganesyan2001,garst2010,metlitski2010,woelfle2007,Lamas2009, Lamas2011}.

The next subleading instability is in the $J=1^-$ channel with $L=1$
and $S=1$ as shown in  Eq. (\ref{eq:landau_0}), which is a spin-current
channel.
The generated order parameters are spin-orbit coupled.
For the channel of $J_z=0$, the generated SO coupling at the
single-particle level exhibits the three-dimensional (3D) Rashba type as
\bea
H_{so,1^-}= |n_z|  \sum_k \psi^\dagger_\alpha(\vec k )
(k_x \sigma_y -k_y \sigma_x)_{\alpha\beta}
\psi_\beta(\vec k),
\label{eq:so1-}
\eea
where $|n_z|$ is the magnitude of the order parameter.
The same result was also obtained recently in Ref. \onlinecite{sogo2011}.
In the absence of spin-orbit coupling, the $L=S=1$ channel
Pomeranchuk instability was studied in Refs. [\onlinecite{wu2004}] and [\onlinecite{wu2007}],
which exhibits the unconventional magnetism with both isotropic
and anisotropic versions.
They are particle-hole channel analogies of the $p$-wave triplet
Cooper pairings of $^3$He isotropic $B$ and anisotropic $A$ phases,
respectively.
In the isotropic unconventional magnetic state, the total angular momentum
of the order parameter is $J=0$, which exhibits the
$\vec k \cdot \vec \sigma$-type spin-orbit coupling.
This spin-orbit coupling is generated from interactions through a phase
transition and, thus, was denoted as the spontaneous  generation of spin-orbit
coupling.
In Eq. (\ref{eq:so1-}), the spin-orbit coupling that appears at the mean-field
single-particle level cannot be denoted as spontaneous
because the magnetic dipolar interaction
possesses the spin-orbit nature.
Interestingly, in the particle-particle channel, the dominant Cooper
pairing channel has the same partial-wave property of
$L=S=J=1$ \cite{li2010}.

The instability in the $J=1^-$ (spin current) channel is weaker than that
in the $1^+$ (ferromagnetism) channel because the magnitude of Landau
parameters is larger in the former case.
The $1^-$ channel instability should occur after the appearance
of ferromagnetism.
Since spin-current instability breaks parity, whereas,
ferromagnetism does not, this transition is a genuine phase transition.
For simplicity, we consider applying an external magnetic field along the
$z$ axis in the ferromagnetic state to remove the spin texture structure.
Even though the $J=1^+$ and $1^-$ channels share the same property under
rotation transformation,
they do not couple at the quadratic level because of their different parity
properties.
The leading-order coupling occurs at the quartic order as
\bea
\delta F= \beta_1 (\vec n \cdot \vec n) (\vec S \cdot \vec S)
+\beta_2 |\vec n \times \vec S|^2,
\eea
where $\vec n$ and $\vec S$ represent the order parameters in the
$J=1^-$ and $1^+$ channels, respectively.
$\beta_1$ needs to be positive to keep the system stable.
The sign of $\beta_2$ determines the relative orientation
between $\vec n$ and $\vec S$.
It cannot be determined purely from the symmetry
analysis but depends on microscopic energetics.
If $\beta_2>0$, it favors $\vec n \parallel \vec S$,
and $\vec n\perp \vec S$ is favored at $\beta_2<0$.


\section{The spin-orbit coupled collective modes}
\label{sect:coll}

In this section, we investigate another important feature of the
Fermi liquid, the collective modes, which again exhibit
the spin-orbit coupled nature.

\subsection{Spin-orbit coupled Boltzmann equation}

We employ the Boltzmann equation to investigate the collective modes
in the Fermi liquid state\cite{negele1988}
\begin{widetext}
\bea
&&\frac{\partial}{\partial t} n(\vec r, \vec k, t)
-\frac{i}{\hbar} [\epsilon(\vec r, \vec k, t),  n(\vec r, \vec k, t)]
+\frac{1}{2}\sum_i
\Big\{ \frac{\partial \epsilon (\vec r, \vec k, t)}{\partial k_i},
\frac{\partial n (\vec r, \vec k, t)}{\partial r_i} \Big\} -
\frac{1}{2} \sum_i
\Big\{ \frac{\partial \epsilon (\vec r, \vec k, t)}{\partial r_i},
\frac{\partial n (\vec r, \vec k, t)}{\partial k_i} \Big\}
=0,
\eea
\end{widetext}
where $n_{\alpha\alpha^\prime}(\vec r, \vec k, t)$ and  $\epsilon_{\alpha\alpha^\prime}
(\vec r, \vec k, t)$ are the density and energy matrices
for the coordinate $(\vec r, \vec k)$ in the phase space and
$[, ]$ and $\{, \}$ mean the commutator and
anticommutator, respectively.
Under small variations in $n_{\alpha\alpha^\prime}(\vec r, \vec k, t)$
and $\epsilon_{\alpha\alpha^\prime}(\vec r, \vec k, t)$,
\bea
n_{\alpha\alpha^\prime}(\vec r, \vec k, t)&=&n_0(k) \delta_{\alpha\alpha^\prime}
+\delta n_{\alpha\alpha^\prime}(\vec r, \vec k, t),\nn \\
\epsilon_{\alpha\alpha^\prime}(\vec r, \vec k, t)&=&\epsilon(k)
\delta_{\alpha\alpha^\prime}+\int \frac{d^3 k^\prime}{(2\pi)^3}
f_{\alpha\alpha^\prime,\beta\beta^\prime}(\hat k, \hat k^\prime) \nn \\
&\times&\delta n_{\beta\beta^\prime} (\hat k^\prime).
\eea
the above Boltzmann equation can be linearized.
Plugging the plane-wave solution of
\bea
\delta n_{\alpha\alpha^\prime}(\vec r, \vec k, t)
=\sum_q \delta n_{\alpha\alpha^\prime}(\vec k) e^{i (\vec q \cdot \vec r-\omega t)},
\eea
we arrive at
\bea
\delta n_{\alpha\alpha^\prime} (\hat k)&-&
\frac{1}{2}\frac{\cos\theta_k}{s-\cos \theta_k} \sum_{\beta\beta^\prime}
\int d\Omega_{k^\prime}
\frac{N_0}{4\pi} f_{\alpha\alpha^\prime, \beta\beta^\prime}(\hat k, \hat k^\prime)
\nn \\
&\times& \delta n_{\beta\beta^\prime}(\hat k^\prime)=0,
\eea
where $s$ is the dimensionless parameter $\omega/(v_f q)$.
The  propagation direction of the wavevector $\vec q$ is defined
along the $z$-direction.

In the spin-orbit decoupled basis defined as $\delta n_{LmSs_z}$ in Sec.
\ref{sect:basis}, the linearized Boltzmann equation becomes
\bea
\delta n_{LmSs_z}&+&\Omega_{LL^\prime; m} (s) F_{L^\prime m^\prime S s_z;
L^{\prime\prime}m^{\prime\prime}S s_z^{\prime\prime}}
\delta n_{L^{\prime\prime}m^{\prime\prime}Ss_z^{\prime\prime}}=0,
\label{eq:boltzm_1} \nn \\
\eea
where $\Omega_{LL^\prime}(s)$ is equivalent to the particle-hole channel
Fermi bubble in the diagrammatic method as
\bea
\Omega_{LL^\prime;m}(s)=-\int d\Omega_{\hat k} Y^*_{Lm}(\hat k)
Y_{L^\prime m }(\hat k) \frac{\cos \theta_k }{s-\cos\theta_k}.
\eea
For later convenience, we present $\Omega_{LL^\prime;m}$ in several
channels of $LL^\prime$ and $m$ as follows
\bea
\Omega_{00;0}(s)&=&1-\frac{s}{2} \ln |\frac{1+s}{1-s}|+ i\frac{\pi}{2} s
\Theta(s<1),\nn \\
\Omega_{10;0}(s)&=&\Omega_{01;0}=\sqrt 3 s \Omega_{00;0}(s), \nn \\
\Omega_{11;0}(s)&=& 1+3s^2 \Omega_{00;0}(s),\nn \\
\Omega_{11;1}(s)&=&\Omega_{11;-1}(s)=-\frac{1}{2}
\Big[1-3(1-s^2) \Omega_{00;0}(s)\Big].
\nn \\
\eea

Equation (\ref{eq:boltzm_1}) can be further simplified by using the
spin-orbit coupled basis $\delta n_{JJ_z;LS}$ defined in Sec. \ref{sect:basis},
\bea
\delta n_{JJ_z;LS}&+&\sum_{J^\prime;LL^\prime} K_{JJ_zLS;J^\prime J_z L^\prime S}(s)
F_{J^\prime J_zL^\prime S; J^\prime J_z L^{\prime\prime} S}\nn \\
&\times& \delta n_{J^\prime J_z L^{\prime\prime} S}=0,
\label{eq:sound_eign1}
\eea
where the matrix kernel $K_{JJ_zLS;J^\prime J_z L^\prime S}$ reads
\bea
K_{JJ_zLS;J^\prime J_z L^\prime S} (s) &=&\sum_{m s_z}
\avg{LmSs_z|JJ_z} \avg{L^\prime m S s_z| J^\prime J_z} \nn \\
&\times& \Omega_{LL^\prime;m}(s).
\eea

\subsection{The spin-orbit coupled sound modes }
Propagating collective modes exist if Landau parameters are positive.
In these collective modes, interactions among quasiparticles
rather than the hydrodynamic collisions provide the restoring force.
Because only the spin channel receives renormalization from the magnetic dipolar
interaction, we only consider spin channel collective modes.
The largest Landau parameter is in the $(1^+001)$ channel in which
the spin oscillates along the direction of $\vec q$.
The mode in this channel is the longitudinal spin zero sound.
On the other hand, due to the spin-orbit coupled nature, the Landau parameters
are negative in the transverse spin channels of $(1^+ \pm1 \,\,\, 0 \,\, \pm1)$,
and thus no propagating collective modes exist in these channels.
The hybridization between $(1^+001)$ and $(1^+021)$ is small as shown in
Eq. (\ref{eq:eigen_1}), and the Landau parameter in the $(1^+021)$ channel
is small, thus, this channel also is neglected below for simplicity.

Because the propagation wave vector $\vec q$ breaks the parity and
3D rotation symmetries, the $(1^+001)$ channel couples to other channels
with the same $J_z$.
As shown in Eq. (\ref{eq:sound_eign1}), the coupling strengths
depend on the magnitudes of Landau parameters.
We truncate Eq. (\ref{eq:sound_eign1}) by keeping the orbital partial-wave channels of $L=0$ and $L=1$ because Landau parameters with
orbital-partial waves $L\ge 2$  are negligible.
There are three channels with $L=S=1$ as $(0^-011)$, $(1^-011)$,
and $(2^-011)$.
We further check the symmetry properties of these four modes under
the reflection with respect to any plane containing  $\vec q$.
The mode of $(1^-011)$ is even and the other three are odd, thus
it does not mix with them.
The Landau parameter in the $(2^-011)$ channel is calculated as
$\frac{\pi}{20}\lambda$, which is $1$ order smaller than those
in $(1^+001)$ and $(1^-001)$, thus this channel is also neglected.
We only keep these two coupled channels $(1^+001)$ and $(1^-001)$
in the study of collective spin excitations.

The solution of the two coupled modes reduces to a $2\times 2$ matrix
 linear equation as
\begin{widetext}
\bea
\left( \begin{array}{cc}
1+\Omega_{00;0}(s) F_{1001;1001}
& s \Omega_{00;0}(s)F_{0011;0011}\\
s \Omega_{00;0}(s)F_{1001;1001} &
1+ \Omega_{00;0}(s) F_{0011;0011}
\end{array}
\right)
\left(\begin{array}{c}
\delta n_{1001}\\
\delta n_{0011}
\end{array}
\right)=0,
\label{eq:eigen_sound}
\eea
where the following relations are used
\bea
K_{1001;1001}(s)&=& \Omega_{00;0}(s)\nn \\
K_{1001;0011}(s)&=& K_{0011;1001}(s)
=\avg{0010|10}\avg{1010|00} \Omega_{01;0}(s)
= s \Omega_{00;0}(s)\nn \\
K_{0011,0011}(s)&=&\sum_m |\avg{1m1-m|00}|^2 \Omega_{11;m}(s)
=\frac{1}{3}\Omega_{11;0}(s) +\frac{2}{3} \Omega_{11;1} (s)
= \Omega_{00;0}(s).
\eea
\end{widetext}
The condition of the existence of nonzero solutions of Eq. (\ref{eq:eigen_sound})
becomes
\bea
(1-s^2) \Omega^2_{00;0}(s)+2 \Omega_{00;0}(s)
\frac{F_+}{F_\times^2}+\frac{1}{F_\times^2}=0,
\label{eq:zerosoundvelocity}
\eea
where $F_+=(F_{1001:1001}+F_{0011;0011})/2$ and $F_{\times}=\sqrt{F_{1001:1001} F_{0011;0011}}$.

\begin{figure}[tbp]
\centering\epsfig{file=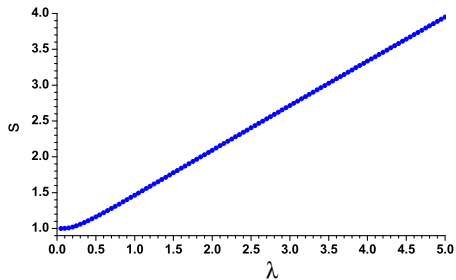,clip=1,width=0.8\linewidth, angle=0}
\caption{(Color online) The sound velocity $s$ in the unit of $v_f$ v.s. the
dipolar coupling strength $\lambda$.
At $0<\lambda\ll 1$, $s(\lambda)\approx 1+0^+$.
On the order of $\lambda\gg 1$, $s(\lambda)$ becomes linear
with the slope indicated in Eq. (\ref{eq:asym_2}).
}
\label{fig:zerosoundvelocity}
\end{figure}

Let us discuss several important analytical properties of its solutions.
In order for collective modes to propagate in Fermi liquids, its
sound velocity must satisfy $s>1$, otherwise it enters the particle-hole
continuum and is damped, a mechanism called Landau damping.
We can solve Eq. (\ref{eq:zerosoundvelocity}) as
\bea
\Omega_{00;0}^{\pm}(s)= \frac{F_+ \pm \sqrt{F_+^2 +(s^2-1)F^2_{\times}}}
{(s^2-1)F^2_{\times}}.
\eea
Only the expression of the $\Omega_{00;0}^{-}(s)$ is consistent with $s>1$
and is kept.
The other branch has no solution of the propagating collective modes.

Let us analytically check two limits with large and small values of
$\lambda$, respectively.
In the case of $0<\lambda\ll 1$ such that $s\rightarrow 1+0^+$,
Eq. (\ref{eq:zerosoundvelocity}) reduces to
\bea
\Omega_{00;0}(s_{\lambda \ll 1})\approx 1-\frac{1}{2} \ln 2 +
\frac{1}{2} \ln (s-1)
= - \frac{1}{2F_+}.
\eea
Its sound velocity solution is
\bea
s_{\lambda \ll 1}&\approx&1+2e^{-2 \big( 1+\frac{1}{2F_+} \big) }
= 1+2e^{-2-\frac{12}{7 \pi \lambda}}.
\label{eq:asym_1}
\eea
The eigenvector can be easily obtained as $\frac{1}{\sqrt{2}}(1,1)^T$,
which is an equal mixing between these two modes.
On the other hand, in the case of $\lambda \gg 1$, we also
expect $s\gg 1$, and thus Eq. (\ref{eq:zerosoundvelocity}) reduces to
\bea
\Omega_{00;0}(s_{\lambda \gg 1})\approx -\frac{1}{s F_\times}
 = -\frac{1}{3s^2},
\eea
whose solution becomes
\bea
s_{\lambda \gg 1}\approx \frac{F_\times}{3}
=\frac{\pi}{3 \sqrt{3}} \lambda.
\label{eq:asym_2}
\eea
In our case,  $F_{1001}$ is larger than $F_{0011}$ but is on the
same order.
The eigenvector can be solved as $\frac{1}{\sqrt{2F+}}(\sqrt {F_{0011}},
\sqrt {F_{1001}})^T$ in which the weight of the $(0011)$ channel
is larger.

\begin{figure}[tbp]
\centering\epsfig{file=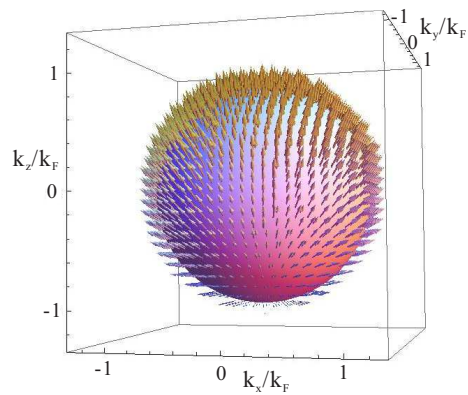,clip=1,width=0.7\linewidth, angle=0}
\caption{(Color online) The spin configuration [Eq. (\ref{eq:spinconfig})]
of the zero-sound mode over the Fermi
surface shows hedgehog-type topology at $\lambda=10$.
The common sign of $u_1$ and $u_2$ is chosen to be positive, which gives rise to
the Pontryagin index $+1$. Although the hedgehog configuration is distorted
in the $z$ component, its topology does not change for any values
of $\lambda$ describing the interaction strength.
}
\label{fig:spinconfig}
\end{figure}

The dispersion of the sound velocity $s$ with respect to the dipolar
interaction strength $\lambda$ is solved numerically as presented in
Fig. \ref{fig:zerosoundvelocity}.
Collective sound excitations exist for all the interaction strengths
with $s>1$.
In both limits of $0\ll\lambda\ll 1$ and $\lambda\gg 1$, the numerical
solutions agree with the above asymptotic analysis
of Eqs. (\ref{eq:asym_1}) and (\ref{eq:asym_2}).
In fact, the linear behavior of $s(\lambda)$ already appears
at $\lambda \sim 1$, and the slope is around $0.6$.
For all the interaction strengths, the  $(1^+001)$ and $(0^-011)$
modes are strongly hybridized.

This mode is an oscillation of spin-orbit coupled Fermi surface distortions.
The configuration of the $(0^-011)$ mode exhibits an oscillating spin-orbit
coupling of the $\vec k \cdot \vec \sigma$ type.
This is the counterpart of the isotropic unconventional magnetism, which
spontaneously generates the $\vec k \cdot \vec \sigma$-type coupling
\cite{wu2004,wu2007}.
The difference is that, here, it is a collective excitation rather than an
instability.
It strongly hybridizes with the longitudinal spin mode.
The spin configuration over the Fermi surface can be represented as
\bea
\vec s(\vec r, \vec k, t) = \left(\begin{array}{c}
u_2 \sin\theta_{\vec k} \cos\phi_{\vec k} \\
u_2 \sin\theta_{\vec k} \sin\phi_{\vec k} \\
u_2 \cos \phi_{\vec k} +u_1
\end{array}
\right) e^{i(\vec q \cdot \vec r -s q v_f t)},
\label{eq:spinconfig}
\eea
where $(u_1, u_2)^T$ is the eigenvector for the collective mode.
We have checked that for all the values of $\lambda$,
$|u_2|>|u_1|$ is satisfied with no change in their relative sign,
thus the spin configuration as shown in Fig. \ref{fig:spinconfig}
is topologically non-trivial with the Pontryagin index $\pm 1$
which periodically flips the sign with time and the spatial coordinate
along the propagating direction.
It can be considered as a topological zero sound.

\section{Conclusions}
\label{sect:conc}

To summairze, we have presented a systematic study on the Fermi liquid
theory with the magnetic dipolar interaction, emphasizing its intrinsic
spin-orbit coupled nature.
Although this spin-orbit coupling does not exhibit at the single-particle
level, it manifests in various interaction properties.
The Landau interaction function is calculated at the Hartree-Fock level
and is diagonalized by the total angular momentum and parity quantum
numbers.
The Pomeranchuk instabilities occur at the strong magnetic dipolar
interaction strength generating effective spin-orbit coupling
in the single-particle spectrum.

We have also investigated novel collective excitations in the
magnetic dipolar Fermi liquid theory.
The Boltzmann transport equations are decoupled in the spin-orbit
coupled channels.
We have found an exotic collective excitation, which
exhibits spin-orbit coupled Fermi surface
oscillations with a topologically nontrivial spin configuration,
which can be considered as a topological zero-sound-like mode.

\acknowledgments
Y. L. and C. W. were supported by the AFOSR YIP program and NSF-DMR-1105945.


\begin{thebibliography}{10}

\bibitem{ospelkaus2008}
S. Ospelkaus, K.~K. Ni, M.~H.~G. {de Miranda}, B. Neyenhuis, D. Wang, S.
  Kotochigova, P.~S. Julienne, D.~S. Jin, and J. Ye,
  Faraday Discuss. {\bf 142}, 351 (2009)..

\bibitem{ni2008}
K.~K. Ni, S. Ospelkaus, M.~H.~G. {de Miranda}, A. Pe'er, B. Neyenhuis, J.~J.
  Zirbel, S. Kotochigova, P.~S. Julienne, D.~S. Jin, and J. Ye, Science {\bf
  322},  231  (2008).

\bibitem{chotia2011}
A. Chotia, B. Neyenhuis, S. A. Moses, B. Yan, J. P. Covey, M. Foss-Feig,
A. M. Rey, D. S. Jin, and J. Ye,
Phys. Rev. Lett. {\bf 108}, 080405 (2012).


\bibitem{sogo2009}
T. Sogo, L. He, T. Miyakawa, S. Yi, H. Lu, and H. Pu, New J. Phys. {\bf 11},
  055017  (2009).

\bibitem{miyakawa2008}
T. Miyakawa, T. Sogo, and H. Pu, Phys. Rev. A {\bf 77},  061603  (2008).


\bibitem{fregoso2009}
B.~M. Fregoso, K. Sun, E. Fradkin, and B.~L. Lev, New J. Phys. {\bf
  11},  103003  (2009).

\bibitem{chan2010}
C. K. Chan, C. Wu, W. C. Lee, S. Das Sarma, Phys. Rev. A {\bf 81}, 023602 (2010),

\bibitem{ronen2009}
S. Ronen and J. L. Bohn, Phys. Rev. A 81, 033601 (2010).

\bibitem{lin2009}
C. Lin, E. Zhao, and W. V. Liu, Phys. Rev. B {\bf 81}, 045115 (2010).


\bibitem{yamaguchi2010}
Y. Yamaguchi, T. Sogo, T. Ito, T. Miyakawa, Phys. Rev. A {\bf 82}, 013643 (2010).

\bibitem{Li2010}
Q. Li, E.~H. Hwang, S. Das Sarma, Phys. Rev. B {\bf 82}, 235126 (2010).

\bibitem{chang2008}
C. M. Chang, W. C. Shen, C. Y. Lai, P. Chen, D. W. Wang,
Phys. Rev. A {\bf 79}, 053630 (2009).

\bibitem{zhang2010}
J. N. Zhang, S. Yi, Phys. Rev. A {\bf 81}, 033617 (2010);
J. N. Zhang, S. Yi, Phys. Rev. A {\bf 80}, 053614 (2009).

\bibitem{sun2010}
K. Sun, C. Wu, S. Das Sarma, Phys. Rev. B {\bf 82}, 075105 (2010).


\bibitem{youn2010}
S. H. Youn, M. Lu, U. Ray, and B. ~L. Lev,
\newblock {\em Phys. Rev. A}{ \bf 82}, 043425 (2010).


\bibitem{lu2010}
M. Lu, S. H. Youn,  and B. L. Lev,
\newblock {\em Phys. Rev. Lett.}{ \bf 104}, 063001 (2010).

\bibitem{lu2011}
M. Lu, N. Q. Burdick, S. H. Youn, B. L. Lev,
Phys. Rev. Lett. {\bf 107}, 190401 (2011).

\bibitem{koch2008}
T. Koch, T. Lahaye, J. Metz, B. Frohlich, A. Griesmaier, and T. Pfau,
Nature Physics {\bf 4},  218  (2008).

\bibitem{lahaye2009}
T. Lahaye, C. Menotti, L. Santos, M. Lewenstein, and T. Pfau,
Rep. Prog. Phys. {\bf 72}, 126401 (2009).


\bibitem{lahaye2009a}
T. Lahaye, J. Metz, T. Koch, B. Frohlich, A. Griesmaier, and T. Pfau,
Pushing the Frontiers of Atomic Physics,  {\bf 1}, 160 (2009).

\bibitem{menotti2007}
C. Menotti, M. Lewenstein, T. Lahaye, and T. Pfau,
in \emph{Dipolar Interaction in Ultra-Cold Atomic Gases},
edited by A. Campa, A. Giansanti, G. Morigi and F. S. Labini,
AIP Conf. Proc. No. 970 (AIP, New York, 2008), p. 332.

\bibitem{quintanilla2009}
J. Quintanilla, S. T. Carr, and J. J. Betouras,
Phys. Rev. A {\bf 79}, 031601 (R) (2009).

\bibitem{fregoso2009a}
B. M. Fregoso, E. Fradkin, Phys. Rev. Lett. {\bf 103}, 205301 (2009)

\bibitem{fregoso2010}
B. M. Fregoso, E. Fradkin, Phys. Rev. B {\bf 81}, 214443 (2010).

\bibitem{you1999}
L. You and M. Marinescu, Phys. Rev. A {\bf 60}, 2324 (1999).

\bibitem{baranov2002}
M.~A. Baranov, M.~S. Mar'enko, V.~S. Rychkov, and G.~V. Shlyapnikov,
  Phys. Rev. A {\bf 66},  013606  (2002).

\bibitem{baranov2004}
M.~A. Baranov, {\L}. Dobrek, and M. Lewenstein, Phys. Rev. Lett. {\bf
  92},  250403  (2004).

\bibitem{baranov2008a}
M.~A. Baranov, Physics Reports {\bf 464},  71  (2008).


\bibitem{bruun2008}
G. M. Bruun, E. Taylor, Phys. Rev. Lett. {\bf 101}, 245301 (2008).

\bibitem{levinsen2011}
J. Levinsen, N. R.  Cooper,  and G. V. Shlyapnikov,
Phys. Rev. A {\bf 84}, 013603 (2011).

\bibitem{potter2010}
A. C. Potter, E. Berg, D. W. Wang, B. I. Halperin, and E.
Demler, Phys. Rev. Lett. {\bf 105}, 220406 (2010).

\bibitem{lutchyn2010}
R. M. Lutchyn, E. Rossi, and S. Das Sarma, Phys. Rev. A {\bf 82}, 061604 (2010).


\bibitem{wu2010}
C. Wu, and J. E. Hirsch, Phys. Rev. B {\bf 81}, 020508 (R) (2010).

\bibitem{samokhin2006}
K. V. Samokhin, and M. S. Mar'enko, Phys. Rev. Lett.  {\bf 97},
197003 (2006).

\bibitem{shi2009}
T. Shi, J. N. Zhang. C. P. Sun and S. Yi,
Phys. Rev. A  {\bf 82}, 033623, (2010).

\bibitem{kain2011}
B. Kain, Hong Y. Ling, Phys. Rev. A {\bf 83}, 061603(R), (2011).


\bibitem{li2010}
Y. Li, C. Wu, Scientific Reports {\bf 2}, 392 (2012).

\bibitem{leggett1975}
A. J.  Leggett,
\newblock {\em Rev. Mod. Phys.}{ \bf 47}, 331 (1975).

\bibitem{volovik2009}
G. E. Volovik, {\it ``The Universe in a Helium Droplet'' },
\newblock {(Oxford University Press, Oxford 2009).}

\bibitem{balian1963}
R. Balian, N. R. Werthamer,
\newblock {\em Phys. Rev.}{ \bf 131}, 1553 (1963).


\bibitem{anderson1961}
P. W. Anderson,  P. Morel, \newblock {\em Phys. Rev.}{ \bf 123}, 1911 (1961).

\bibitem{brinkman1974}
W. F. Brinkman, J. W. Serene, and P. W. Anderson,
\newblock {\em Phys. Rev. A}{ \bf 10}, 2386 (1974).


\bibitem{negele1988}
W. Negele and H. Orland, {\it Quantum Many-Particle Systems},
(Perseus Books, New York, 1988).

\bibitem{sogo2011}
T. Sogo, M. Urban, P. Schuck, T. Miyakawa,
Phys. Rev. A 85, 031601(R) (2012).

\bibitem{pomeranchuk1959}
I. J. Pomeranchuk, Sov. Phys. JETP {\bf 8}, 361 (1959).

\bibitem{oganesyan2001}
V. Oganesyan, S. A. Kivelson, and E. Fradkin, Phys. Rev. B {\bf 64},
195109 (2001).

\bibitem {garst2010}
M. Garst and A. V. Chubukov, Phys. Rev. B {\bf 81}, 235105 (2010).

\bibitem{metlitski2010}
M. A. Metlitski and S. Sachdev, Phys. Rev. B {\bf 82}, 075127 (2010)

\bibitem{woelfle2007}  P. W{\"o}lfle, and A. Rosch,
Journal of Low Temperature Physics, {\bf 147}, 165 (2007).

\bibitem{Lamas2009}  C. A. Lamas, D. C. Cabra, and N. Grandi,
Phys. Rev. B {\bf 80}, 075108 (2009).

\bibitem{Lamas2011}  C. A. Lamas, D. C. Cabra, and N. E. Grandi,
Int. J. of Mod. Phys. B {\bf 25}, 3539 (2011).

\bibitem{wu2004}
C. Wu, and S. C. Zhang, Phys. Rev. Lett. {\bf 93}, 036403(2004).

\bibitem{wu2007}
C. Wu, K. Sun, E. Fradkin, and S. C. Zhang, Phys. Rev. B {\bf 75}, 115103 (2007).










\end{thebibliography}
\end{document}